# Crack formation in wet colloidal pillars

By *Justin Beroz\*, Alvin T. L. Tan\*, Ken Kamrin, A. John Hart*

Department of Mechanical Engineering
Massachusetts institute of Technology
77 Massachusetts Avenue
Cambridge, MA 02139 USA
E-mail: (jberoz@mit.edu)
\*these authors contributed equally

Abstract:

We investigate the initiation of cracks in vertically freestanding water-saturated colloidal pillars constructed using a direct-write technique. Paradoxically, the cracks form during drying at the free end, far from the substrate, where the particle network is unconstrained in contracting its volume as it bears compression by a uniform capillary pressure acting at its outer surface. This is explained by a dominant balance of wetting energy terms, from which follows a simple relationship between the particle size and pillar dimensions that captures the presence or absence of cracks. This relationship provides a practical guideline for fabricating crack-free colloidal structures.

The propagation of a crack in an elastic material is energetically favorable when the strain energy released exceeds the corresponding increase of surface energy. This is the basis of elastic fracture mechanics used to predict material failure in innumerable scenarios [1–3]. Clearly, crack propagation in a homogenous material subject to isotropic compression is never energetically favorable because the introduction of a crack would simply create surface energy without relieving any bulk stress. One might suppose that the slow drying of a liquid-saturated colloidal solid with an unconstrained boundary is an example of such a case, wherein a capillary pressure at the outer surface of the network of colloidal particles causes isotropic compression and volumetric shrinkage. Indeed, the established modelling treatment is as an elastic continuum material that shrinks volumetrically during drying in a manner analogous to thermal contraction [4]. In drying colloidal films, a fixed constraint with the underlying substrate is said to be responsible for generating tensile stress in the film, causing cracks [5–7]; the prevention of cracks is technologically important for fabrication of photonic crystals [8,9], as well as solidification of paints, inks and coatings [10–12].

We constructed a set of vertically freestanding solid colloidal pillars by controlled dispensing of a colloidal suspension from a needle onto a substrate mounted on a translational stage (Fig. 1a, [13]). In a subset of the pillars, we observed the formation of cracks during drying either throughout their height or only at the top, i.e., far away from the fixed constraint at the substrate and thus contradictory to the established approach. In this paper, we detail our experimental findings and derive a simple geometric criterion for crack initiation that is based on a dominant balance of wetting energy terms; a departure from the typical balance of strain and surface energies for dry materials.

The colloidal structures are built in a uniform temperature environment ($70^0$ C) by positioning the needle orifice near the substrate and dispensing the particle suspension to establish a liquid bridge, which is then maintained by continuous dispensing to balance the rate of water evaporation (~$10^{-2}$ μl/s) (Fig. 1a). Particles accumulate into a solid layer at the base of the liquid bridge and the substrate is retracted at a controlled rate equaling the increase in layer thickness (~$10^0$ μm/s); this maintains the liquid bridge while precipitating a solid particle structure from its bottom which can reach aspect ratios >10. The structure precipitates wet, i.e., saturated with water between the particles, and eventually establishes a steady state build rate with a drying front at a fixed distance $L$ below the liquid bridge (Fig. 1b). Construction is terminated by halting dispensing, which collapses the liquid bridge and causes the remaining liquid in the structure to evaporate (Fig. 1c). In all experiments, the suspension dispensed through the needle is monodisperse polystyrene spheres in deionized water, with radii $a \in [44, 5000]$ nm at volume fraction $\phi_1 \approx 0.025$.

We find that the subset of structures that develop cracks can be categorized as one of two types: those with shallow cracks in arbitrary orientations, and those with wide and circumferential cracks. The wide circumferential cracks are typically visible *in situ* with video microscope cameras (Fig. 1c-e and Supplementary Video 1), and therefore are the subject of our investigation. Measurements of the radial change of thin cross-sections of the structures over the course of drying show that the capillary pressure causes radial constriction and then expansion to a final radius $R$ upon passage through the drying front (Fig. 1e). These circumferential cracks consistently appear in the particular section of the structure that is wet with its radius constricted to $\approx R$, and we observe this is two instances: (1) in the portion of wet section $l$ just above the drying front during construction (Fig. 1e(i)), and (2) throughout the entire wet section $l$ at the top of the structure during evaporation of the remaining liquid after construction is terminated (Fig. 1e(ii)). In the latter, the radial change and transition through the drying front are approximately uniform across the section $l$ at the top, and in some structures the cracks only appear here (Fig. 1c,d). We have confirmed by electron and x-ray imaging that the particles are close-packed at the surface and throughout the volume [13]. This indicates that initially where the structure's radius is $> R$ there is space between the particles, which we estimate is on the order of a few percent of their radii $a$, and where structure's radius is $< R$ the structure is elastically compressed. The particle packing is typically polycrystalline with order ranging $\sim 10^{0-2} a$, and the location and orientation of the crystalline patches bears no recognizable correlation to that of the cracks (Fig. 2a,b).

At the moment the cracks occur, we estimate that the force exerted on the particles from the capillary pressure $F_\gamma \sim 2\pi\gamma a$ is primarily borne by particle-particle contacts because other forces experienced by the particles – namely viscous $F_\mu \sim 6\pi\mu U a$, gravitational $F_g = \frac{4\pi}{3}\rho_\Delta g a^3$, electrostatic $F_\zeta \sim \pi\varepsilon\zeta^2$, Van der Waals $F_A \sim \frac{Aa}{12\delta^2}$, and thermal $F_T \sim \frac{3\phi_2 k_B T}{4a}$ – are smaller by at least an order of magnitude. Estimates of the quantities in the above approximate formulas are: water surface tension $\gamma \approx 65$ mN/m (at 70 C); dynamic viscosity $\mu \approx 0.4$ mPa.s; water velocity $U \sim 10^{-3}$ mm/s (estimated from dispensing rates); density difference $\rho_\Delta \equiv \rho_{polystyrene} - \rho_{water} \approx 40$ kg/m$^3$;

gravitational acceleration $g = 9.8$ m/s$^2$; dielectric constant of water $\varepsilon \approx 63\varepsilon_0$ (at 70 C); zeta potential $\zeta \approx 50$ mV (measured at room temperature); Hamaker constant $A \approx 1.4 \times 10^{-20}$ J; molecular space between contacting particles $\delta \sim 10^{-1}$ nm [14]; particle volume fraction in the structure $\phi_2 \approx 0.7$ (measured [13]); temperature $T = 70$ C, and Boltzmann constant $k_B = 1.38 \times 10^{-23}$ J/K.

Because the radius of the section is $\approx R$ at the moment the cracks appear, as a lowest order approximation we assume the structure here is negligibly strained and ignore the elastic strain energy in the particles as if they are hard spheres. This leaves only energy quantities related to the liquid and wetting of the particles, and we proceed by considering the cylindrical section of the structure with radius $\approx R$ and height $l$, as depicted in Fig. 1e, just before it initiates a crack. At this order of approximation, the section comprises contacting particles saturated in stationary water at uniform pressure difference $P$ from atmosphere due to the microscopic water menisci wetting the particles at the section's outer surface; the surface energies $\gamma$, $\gamma_{ap}$, $\gamma_{lp}$ correspond respectively to surface areas $A$, $A_{ap}$, $A_{lp}$. Evaporation is treated simply as the removal of liquid at constant temperature.

In a differential time interval, a differential volume of water $dV$ evaporated from the section's outer surface in turn may deform and recede the menisci between the particles so that the differential change in free energy for the section is $dF = -PdV + \gamma dA + \gamma_{lp} dA_{lp} + \gamma_{ap} dA_{ap} = -PdV + \gamma\left(dA + \cos\theta dA_{ap}\right)$. This is simplified using Young's law $\gamma_{ap} - \gamma_{lp} = \gamma\cos\theta$ which defines the contact angle $\theta$, and recognizing that $dA_{ap} = -dA_{lp} > 0$ because the water may recede to expose more air-particle surface at the expense of liquid-particle surface. If we suppose that the particles are immobile, then differential changes in the amount of liquid in the cylindrical section perturb the menisci between the particles from a point of equilibrium, i.e., $dF = -PdV + \gamma\left(dA + \cos\theta dA_{ap}\right) = 0$. Here, the interpretation is that the energy expelled with the evaporated liquid volume $dV < 0$ at pressure difference from atmosphere $P < 0$, i.e., $-PdV < 0$, is equal to the energy acquired creating surface area by deforming and receding the menisci between the particles.

The following scaling of terms recasts $dF$ into a simpler expression. The cylindrical section has a surface area $A_l = 2\pi R l$ and contains a total number of particles $N \sim A_l/a^2$. Changes $dV$, $dA$, $dA_{ap}$ occur concurrently in the differential window of time, so these quantities can be parameterized by the same variable, say $\xi$. The pressure difference scales as $P \sim -\gamma/a$, and around each particle on the section's surface we may ascribe an evaporated volume $dV_a \sim a^3 d\xi$ and changes in surface area quantities $dA_a \sim a^2 d\xi$. This gives $dV = NdV_a \sim A_l a d\xi$ and $dA, dA_{ap} \sim NdA_a \sim A_l d\xi$. Therefore $dF = \left(-c_1 \gamma A_l + \gamma c_2' A_l + c_3' \gamma \cos\theta A_l\right) d\xi = \left(-c_1 \gamma A_l + \gamma c_2 A_l\right) d\xi$ by introducing the appropriate positive coefficients $c_i$ which are functions of geometry and $\theta$, and collecting the surface energy coefficients as $c_2 = c_2' + c_3' \cos\theta$. For $dF = 0$, we simply have $0 = -c_1 \gamma A_l + c_2 \gamma A_l$ and therefore $c_1/c_2 = 1$ regardless of the details of the particular parameterization $\xi$.

The particles cannot, in actuality, all be immobile because some must separate to form a crack. If the particles have a collective mobility such that within the differential time interval they may rearrange to occupy a slightly smaller overall volume – a reasonable supposition based on Fig. 1e – then $dF < 0$ because not all the energy expelled in the evaporated volume is consumed as surface energy in deforming and receding the water menisci. Here $-c_1 \gamma A_l + c_2 \gamma A_l \leq 0$ and therefore $c_1/c_2 \geq 1$, where the ratio $c_1/c_2$ may be viewed as a metric indicating the amount of particle mobility in the sense just described.

For initiation of a crack to be energetically favorable, the same inequality must be satisfied including an additional consumption of surface energy on a characteristic area $A_\delta$ associated with initiating the crack; this gives $-c_1 \gamma A_l + c_2 \gamma A_l + c_\delta \gamma A_\delta \leq 0$, where the factor $c_\delta$ on the crack initiation term is similar to $c_2$. It must be that $A_\delta \sim a^2$ because a crack can only initiate if the section can at least overcome an energy barrier on the order of bringing a particle to or from its outer surface. Rewriting the inequality as $c_2 \left(c_1/c_2 - 1\right)/c_\delta \geq A_\delta/A_l$, evidently for a given particle size cracks will appear if the surface area of the section $A_l$ is made large enough and only for immobile particles, i.e. $c_1/c_2 = 1$, is cracking impossible. Conversely, for a given particle size the section may be crack free provided its surface area is sufficiently small. Succinctly, the section cracks when satisfying

$$g \geq \frac{a^2}{Rl}. \tag{1}$$

The parameter $g$ depends on contact angle $\theta$, particle mobility $c_1/c_2$ and geometry, where $g \geq 0$ and $g = 0$ only for immobile particles. In this way, the section can initiate a crack despite being compressed by a uniform capillary pressure. The result is a ratio of length scales due to having only considered the dominant wetting energy terms.

To compare to experiment, recall again that the height $l$ corresponds to the wet section of the particle structure with its radius constricted to $\approx R$. Clearly, $l$ is smaller during construction than at termination of build (Fig. 1e), which, according to Eq. 1, means that cracks are most likely to initiate at the top of the structure and explains Fig. 1d. For all structures, the quantity $a^2/Rl \approx 10^{-7}$ demarcates those with and without cracks of any kind after drying (Fig. 2c); here we use the maximum values for $l$, i.e., at the top during termination of construction (Fig. 1e(ii)) [15]. For the subset of structures with wide circumferential cracks, the sections in between the cracks have areas $A_H \sim RH$ (Fig. 2a) which approximately saturate Eq. 1, and we find $g \approx 3 \times 10^{-7}$ (Fig. 2d and Fig. 2c, dashed line). These measurements of $H$ were taken only from the top parts of the structures where the cracks formed at termination throughout the wet sections $l \gg H$ (Fig. 1e(ii)), which ensures an approximate upper estimate for $g$. The spacing of the cracks changes in accordance with Eq. 1 as illustrated by Fig. 2a(i) and Fig. 2a(ii).

The fact that we can identify an approximately constant value for $g$ indicates that the particle mobility $c_1/c_2$ is approximately constant at the moment of crack initiation, and $g \ll 1$ presumably indicates $c_1/c_2$ is nearly unity which agrees with our picture of the particles just contacting one another in their close-packed arrangement when the section's radius is $\approx R$. Interestingly, there is a transition between shallower cracks in arbitrary orientations (Fig. 2c, grey circles) and wide circumferential cracks (Fig. 2c, white circles) at $a^2/Rl \approx 2 \times 10^{-8}$. The details and mechanism of crack propagation are of course beyond our free energy argument for crack initiation, and presumably involve energy contributions from additional interparticle forces neglected in our simple picture. Therefore, the reason why we observe the two categories of cracks in the structures is unanswered by our model. Nevertheless, considering only the dominant wetting energy terms is sufficient to develop a simple intuition and criterion for crack initiation that agrees with experiment. We should expect Eq. 1 to be valid for other material combinations of the colloid particles and liquid provided wetting energy dominates at the instant of crack initiation, however the value for $g$ will in general be different. Moreover, $a^2/Rl$ is essentially the ratio of the particle's surface area over the surface area of the wet section of the structure; by replacing $a^2/Rl$ with this surface area ratio, the validity of Eq. 1 may extend to particles and structures of different shapes.

**Acknowledgements**

We thank Prof. John Bush for discussions regarding crack initiation. J.B. was supported by a Department of Defense National Defense Science and Engineering Graduate Fellowship, as well as the Assistant Secretary of Defense for Research and Engineering (Air Force Contract No. FA8721-05-C-0002). A.T.L.T. was supported by a postgraduate fellowship from the Singapore Defence Science Organization. Funding for hardware and experiments was provided by a National Science Foundation CAREER Award to A.J.H. (CMMI-1346638).

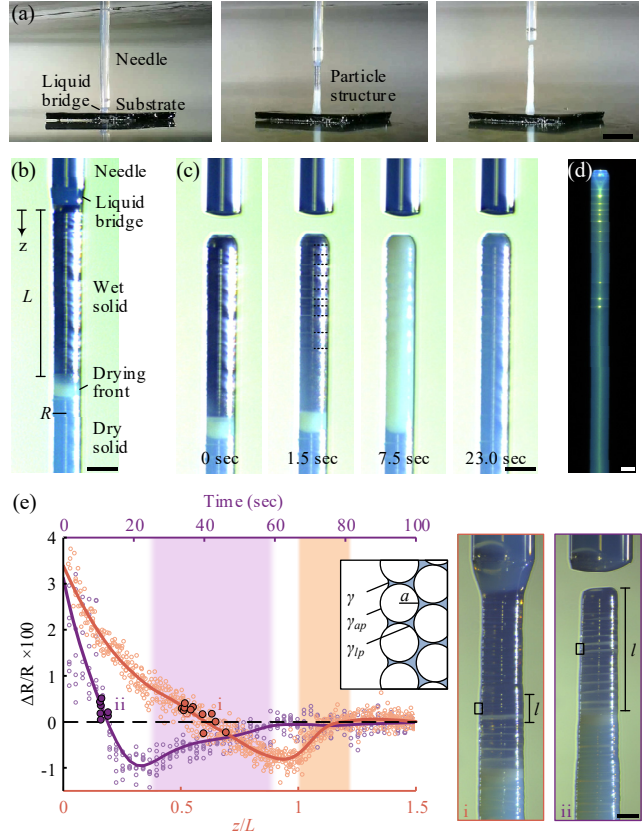

FIG. 1 (color online). (a) Particle structures precipitate from suspension at the base of a liquid bridge at the orifice of a dispensing needle: (left to right) the liquid bridge is first established with the substrate, then withdrawn as particles accumulate, and finally dispensing is halted to terminate construction. (b) The structure precipitates wet, and evaporation occurs from the surface of the wet section, which drives an influx of liquid and particles through the liquid bridge. (c) Shortly after terminating construction, the remaining wet section uniformly develops cracks (black dashed lines) and then transitions through the drying front (opaque white) to become completely dry; (d) this particular structure only developed cracks at the top, which show as bright bands of scattered light under illumination ($a$ = 44 nm). (e) The radial change of a thin cross-section layer of the structure $\Delta R$, relative to its final radius $R$ when dry, exhibits the same contraction and expansion behavior in two cases: (i) during construction, as a function of the layer's distance $z$ below the bottom of the liquid bridge; (ii) at termination, as a function of time. Cracks appear in the sections labeled by $l$, which for both curves correspond to $\Delta R \approx 0$ (solid markers) ahead of the drying front (shaded regions); the data points are an aggregate of 10 layers that each develop a crack and the solid lines are moving averages. All measurements are from the same structure ($a$ = 44 nm) and the results are representative for all structures we built. Scale bars represent 2 mm (a); 200 μm (b-d); 500 μm (e).

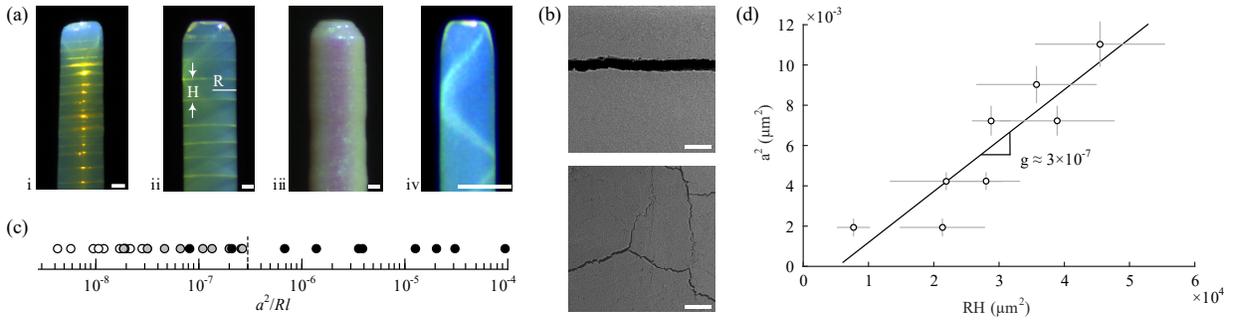

FIG. 2 (color online). (a) Optical images of the top section of particle structures exhibiting (i, ii) wide circumferential cracks, (iii) shallower cracks in arbitrary orientations, and (iv) no cracks, where the structure internally reflects the illumination incident from the top right; (b) exemplary SEM images of a wide circumferential crack (top) and shallower arbitrarily oriented cracks (bottom). (c) $a^2/Rl$ (Eq. 1) demarcates structures with circumferential cracks (white markers), arbitrary oriented cracks (grey markers), and no cracks (black markers). (d) Measurements of the spacing $H$ between circumferential cracks provides an upper estimate on $a^2/Rl$ for the onset of cracking, noted by the dashed line in (c). Scale bars represent 100 μm (a(i)), 50 μm (a(ii-iv)); 5 μm (b).

11